\documentclass[11pt,a4]{article}

\usepackage{amsfonts,amsmath,latexsym}

\title{\bf On Computational Power of Quantum Branching Programs\vspace{0,4cm}\\}

\author{ Farid Ablayev\thanks{Max-Planck Institute for Mathematics, Bonn. Visiting from University of Kazan. 
Supported in part by Russia Fund for Basic Research
99-01-00163 and Fund "Russia Universities" 04.01.52.
{\tt ablayev@ksu.ru }} 
 \and
 Aida Gainutdinova\thanks{Dept. of Computer Science, University of Kazan.
Supported in part by Russia Fund for Basic Research
99-01-00163 and Fund "Russia Universities" 04.01.52.{\tt aida@ksu.ru}} 
 \and
 Marek Karpinski\thanks{Dept. of Computer Science, University of Bonn, and Mathematical Sciences Research Institute, Berkeley.
Supported in part by DFG grants, DIMACS, and IST grant 14036
(RAND-APX). Research at MSRI was supported by NSF grant DMS-9810361. {\tt marek@cs.uni-bonn.de}}  }

\newtheorem{mydef}{Definition}
\newcommand{\bedef}{\begin{mydef}}
\newcommand{\eedef}{\end{mydef}}

\newtheorem{theorem}{Theorem}
\newcommand{\bethm}{\begin{theorem}}
\newcommand{\eethm}{\end{theorem}}

\newtheorem{cors}{Corollary}
\newcommand{\becor}{\begin{cors}}
\newcommand{\eecor}{\end{cors}}

\newtheorem{lemm}{Lemma}
\newcommand{\belem}{\begin{lemm}}
\newcommand{\eelem}{\end{lemm}}

\newtheorem{rft}{Theorem}[section]
\newcommand{\bethmsec}{\begin{rft}}
\newcommand{\eethmsec}{\end{rft}}

\newtheorem{fpr}{Property}
\newcommand{\beprop}{\begin{fpr}}
\newcommand{\eeprop}{\end{fpr}}

\newcommand{\qbp}{\mathrm{QBP}}
\newcommand{\qbps}{\mathrm{QBPs}}
\newcommand{\bp}{\mathrm{BP}}
\newcommand{\bps}{\mathrm{BPs}}

\newcommand{\ra}{\rangle}
\newcommand{\qs}[1]{\mbox{${|\!\!~#1~\!\!\rangle}$}}
\newcommand{\ket}[1]{\mbox{${|\!\!~#1~\!\!\rangle}$}}

\newcommand{\ceiling}[1]{\mbox{$\lceil\!\!~#1~\!\!\rceil$}}

\newcommand{\Endproof}{\hfill$\Box$\\}

\date{}

\begin{document}

\maketitle

\begin{abstract}

 In this paper we study a model of Quantum Branching Program 
($\qbp$) and investigate its computational power. We define several natural
restrictions of a general $\qbp$ model, such as a read-once and
a read-$k$-times $\qbp$, noting that {\em obliviousness} is inherent in
a quantum nature of such programs.

In particular we show that any Boolean function can be computed
deterministically ({\em exactly}) by a read-once $\qbp$ in width $O(2^n)$,
contrary to the analogous situation for quantum finite
automata. Further we display certain symmetric Boolean function which
is computable by a read-once $\qbp$ with $O(\log n)$ width, which requires a 
width $\Omega(n)$ on any deterministic read-once $\bp$ and (classical)
randomized read-once $\bp$ with permanent transitions at each level.

We present a general lower bound for the width of read-once $\qbps$,
showing that the upper bound for a considered symmetric function is
almost tight.

\end{abstract}

\section{Introduction}

Richard Feynman observed in 1982 (\cite{fe82}) that certain quantum
mechanical effects cannot be simulated effectively on a classical
computer. This observation led to a general idea that perhaps
computation based on quantum effects could be much more efficient than the classic one. 
During the last decade the area of research of quantum computation  
was an intensively growing area. Shor's
quantum algorithm for factoring integers \cite{sh94} that runs in polynomial
time is well known now.

As mentioned in \cite{amfr98} quantum computers may consist of two
parts: a quantum part and a classical part with communication between
them. In such a case, a quantum part could be considerably
more expensive than the classical part. Therefore, it might be useful
to construct a quantum part as simple as possible. This motivates a 
study of restricted models of quantum computation.

During the last decade many different restricted quantum computation models have
been investigated.
In particular 
the quantum analogs of Boolean circuits{\bf $NC$} and {\bf $ACC$}
\cite{moni98,grhopo00} has been introduced.  Another model --- quantum finite automata has
been introduced and first studied by Kondacs and Watrous
\cite{kowa97}, see \cite{brpi99} for more information on the subject.
It has been shown that uniform one-way quantum finite automata with
bounded error probability cannot accept all regular languages
\cite{kowa97}. But Ambainis and Freivalds \cite{amfr98} presented
a regular language which can be computed by a quantum finite
automaton with bounded error probability of exponentially smaller size than
the corresponding classical randomized finite automaton. One of the more recent papers comparing classical and restricted quantum computation models is \cite{tnwa01}.

A classical branching program ($\bp$)(see, e.g., \cite{we00}) is a 
convenient nonuniform model for studying various restricted  
variants of computational models. Leveled oblivious permutation
$\bps$ are well known in the complexity theory, their computational power
is remarkable (deterministic leveled oblivious permutation $\bps$ with constant
width have the same power as $\log n$ depth circuits
\cite{ba89}). It seems also that the branching programs are well suited for comparing restricted quantum models with their classical counterparts.

Nakanishi, Hamaguchi, and Kashiwabara introduced also in \cite{nahaka00}
a variant of quantum model of $\bps$ as an extension of
probabilistic $\bps$. In this paper we introduce a model of quantum
$\bps$ different from that of \cite{nahaka00}.  Without loss
of generality we consider leveled $\bps$. For a leveled $\bp$ $P$, all the
paths in a $\bp$ are of the same length, and one can move only from the nodes of the $i$-th
level to the nodes of the $i+1$-th level. We denote by 
$w(P)$ a width of $P$. That is, $w(P)$ is a maximum number of nodes on various
levels of $P$. In our case a superposition of $P$ is any element
$\qs{\psi}=\sum_{i=1}^{w(P)} z_i\qs{q_i}$. We notice that we may need much  
less quantum bits than in a model of \cite{nahaka00}.

In ths paper we investigate a restricted
computational variant of a quantum branching program--- a quantum read-once $\bp$.  First we show that a 
read-once (exact) quantum $\bps$ (noting that the {\em obliviousness} is
inherent in a quantum nature of such programs) can compute an arbitrary
Boolean function.  Next we display a certain symmetric Boolean function
which is computable by a read-once $\qbp$ with $O(\log n)$ width, which
requires a width $\Omega(n)$ on any deterministic read-once $\bp$ and on any 
(classical) randomized read-once $\bp$ with permanent transitions at each
level.  We present a general lower bound for the width of read-once
$\qbps$, showing that the upper bound for a considered symmetric
function is almost tight.

\section{Preliminaries and Definitions}

We consider a $d$-dimensional Hilbert complex space ${\cal H}^d$ with a
norm $||.||$.  That is, for $z\in {\cal H}^d$,
$z=\{z_1,\dots,z_d\}$, $||z||=\sqrt{\sum_{i=1}^d |z_i|^2}$.   

Recall some basic notations from the quantum mechanics. A pure quantum
state (or superposition) of a quantum system QS with $d$ stable states
$\{1,\dots,d\}$ ($d$-state QS) can be expressed by associating an
amplitude $z_i$ (a complex number) to each state $i$ of a QS. Quantum
mechanics uses for that the following notations. Consider the quantum {\em
basis states} $\{|1\rangle,\dots,|d\rangle\}$ where $\{\qs{i}\}$ is
the set of $d$-dimensional orthogonal basis vectors of ${\cal H}^d$
where $\qs{i}$ denotes the unit vector with the value 1 at $i$ and 0
elsewhere. A pure quantum state or a {\em configuration} of a QS can be
written as

\[ |\psi\ra =\sum_{i=1}^dz_i|i\rangle \]
 
or just $\qs{\psi}=(z_1,\dots,z_d)$ where $\qs{\psi}\in{\cal H}^d$.  The
specific notation $\qs{\psi}$ corresponds to the Dirac 'ket'-notation for
a column-vector $(z_1,\dots,z_d)$.  An element $z_i$ of
$\qs{\psi}$ is called an amplitude of the basis state $\qs{i}$ of a QS, and
$|z_i|^2$ is the probability of finding a QS in the state $i$ when QS is
beeing measured.

A time evolution of configurations of a QS in discrete steps is reversible
and conveniently expressed using Heisenberg's matrix mechanics.  
That is, if in a current step a configuration of a QS is $\qs{\psi}$,
then in the next step a configuration of a QS would be $\qs{\psi'}$
where $\qs{\psi'}=U\qs{\psi}$ and $U$ is a $d\times d$ unitary matrix.

We are going to define now a {\em quantum transformation} as follows.  Let
$X=\{x_1,\dots,x_n\}$ be a set of Boolean variables.  Define quantum
transformation ($d$-dimensional quantum transformation) on
$\qs{\psi}\in{\cal H}^d$ as a triple $\langle j,U(0),U(1)\rangle$ where
$j$ is the index of a variable $x_j\in X$, and $U(0),U(1)$ are reversible
transformations of ${\cal H}^d$ presented by unitary $d\times d$
matrices. A quantum transformation $\langle j,U(0),U(1)\rangle$ of
$\qs{\psi}$ acts as follows: $U\qs{\psi}=\qs{\psi'}$.  If $x_j=1$
then $U=U(1)$ else  $U=U(0)$.

\subsection{Definition of a $\qbp$}

A Quantum Branching Program of width $d$
and length $l$ ($(d,l)$-$\qbp$) based on a QS is defined by a triple 

\[ P=\langle T,\qs{\psi_0},F\rangle\] 

where $T$ is a sequence (of length $l$) of $d$-dimensional quantum
transformations of a $d$-state QS:

\[T=(\langle j_i, U_i(0), U_i(1) \rangle)_{i=1}^l,\]

with $\qs{\psi(0)}$ an initial configuration of $P$, and $F\subseteq
 \{1,\dots,d\}$ a set of accepting states.

We define a computation of $P$ over an input
$\sigma=\sigma_1,\dots,\sigma_n\in\{0,1\}^n$ as follows:

\begin{enumerate}
 \item  Computation of $P$ starts from the superposition
$\qs{\psi_0}$. On the $i$-th step, $1\leq i\leq
l$, $P$ transforms a superposition $\qs{\psi}$ to a
superposition
$\qs{\psi'}=U_i(\sigma_{j_i})\qs{\psi}$.

\item After the $l$-th (last) step of a quantum transformation, $P$
measures its configuration $\qs{\psi_\sigma}$ where
$\qs{\psi_\sigma}=U_l(\sigma_{i_l})U_{l-1}(\sigma_{i_{l-1}}) \dots
U_1(\sigma_{i_1})\qs{\psi_0}$. The measurement is presented by a diagonal
zero-one projection matrix $M$ where $M_{ii}=1$ if $i\in F$ and
$M_{ii}=0$ if $i\not\in F$. The probability $p_{accept}(\sigma)$ of
$P$ accepting an input $\sigma$ is defined by

\[ p_{accept}(\sigma)=||M\qs{\psi_\sigma}||^2. \]
 \end{enumerate}

%Note that all computation steps of $P$ are linear except the last
%measurement step. 
 
We call a $\qbp$ $P$ read-once, if  each variable
$x\in\{x_1,\dots,x_n\}$ occurs in a sequence $T$ of quantum
transformations of $P$ at most once.

\subsection{Computation of a Boolean Function}

\begin{itemize}
  \item A $\qbp$ $P$ is said to compute (with an unbounded error)
a Boolean function $f_n:\{0,1\}^n\to \{0,1\}$ if for all $\sigma\in
f^{-1}(1)$ the probability of $P$ accepting $\sigma$ is greater than
1/2 and for all $\sigma\in f^{-1}(0)$ the probability of $P$ accepting
$\sigma$ is at most 1/2.

 \item A $\qbp$ $P$ computes a Boolean function $f_n$ with a bounded error if there exists an
$\varepsilon>0$ such that for all $\sigma\in f^{-1}(1)$ the
probability of $P$ accepting $\sigma$ is at least $1/2+\varepsilon$
and for all $\sigma\in f^{-1}(0)$ the probability of $P$ accepting
$\sigma$ is at most $1/2-\varepsilon$. We call $\varepsilon$ a 
margin, and say that $P$ $(1/2+\varepsilon)$-computes $f_n$.

 \item We say that a $\qbp$ $P$ exactly computes $f_n$ if $P$ computes $f_n$
with the margin 1/2 (with the zero error probability).

\end{itemize}

\section{Computational Properties}

The following property of a simulation of $\qbps$ with complex valued
amplitudes by $\qbps$ with real valued amplitudes is similar to
simulations by quantum Turing machines cf. \cite{beva97}.

\beprop 

Let a $(d,l)$-$\qbp$ $P$ with complex valued amplitudes computes (with an 
 unbouded error, with a bounded error, exactly) a function $f_n$. Then,
 there exists a $(2d,l)$-$\qbp$ $P'$ with real amplitudes within the
 interval $[-1,+1]$ that computes (with an unbouded error, with a bounded
 error, exactly, respectively) the same function $f_n$.

\eeprop

{\bf Proof:} Let us consider a product $U\ket{\psi}$ of a complex
valued $d\times d$ matrix $U$

\[ U= 
     \left( \begin{array}{ccc}
 z_{1,1} &\dots  & z_{1,d} \\
 \vdots &  & \vdots \\
 z_{d,1} &\dots  & z_{d,d} \\
      \end{array}
 \right) \]

where  $z_{i,j}= a_{i,j}+\sqrt{-1}b_{i,j}$,\\ 
with a $d$-dimensional complex valued vector $\ket{\psi}$

%\[ \ket{\psi} =(z_1,\dots,z_d)^T\]

\[ \ket{\psi}= 
     \left( \begin{array}{c}
 z_{1} \\
 \vdots \\
 z_{d}\\
      \end{array}
 \right) \]

where $z_i=a+\sqrt{-1}b_i$. \\
The above product can be simulated by a product of a 
$A\ket{v}$ of $2d\times 2d$ real valued matrix $A$ 

 \[ A= 
     \left( \begin{array}{ccc}
 Z_{1,1} &\dots  & Z_{1,d} \\
 \vdots &   & \vdots \\
 Z_{d,1} &\dots  & Z_{d,d} \\
      \end{array}
 \right) \]

where $Z_{i,j} =\left(\begin{array}{cc}
 a_{i,j} & b_{i,j} \\
-b_{i,j} & a_{i,j} \\
\end{array}
\right)$ is matrix presentation of the complex number $z_{i,j}$, and \\ 
$\ket{v}$ is a $2d$-dimensional real valued vector

\[ \ket{v}= 
     \left( \begin{array}{c}
 a_{1} \\
 b_1 \\
 \vdots \\
 a_{d}\\
 b_d
      \end{array}
 \right) \]

We notice that all amplitudes $z$ in superpositions
$\ket{\psi}$ of $P$ are such that the real $a$ and imaginary $b$ part of
$z=a+\sqrt{-1}b$ are from intervals $[-1,+1]$.  From the above it is easy to  
conclude that a $d$-dimensional $l$-length complex valued quantum
transformations of a $d$-state QS

\[(\langle j_i, U_i(0), U_i(1) \rangle)_{i=1}^l,\]

can be simulated by the corresponding $2d$-dimensional $l$-length real
valued quantum transformations of a $2d$-state QS

\[(\langle j_i, A_i(0), A_i(1) \rangle)_{i=1}^l.\]

 \Endproof

Below we show that a bounded error read-once $\qbp$s are powerful enough
to compute an arbitrary Boolean function. By contrast, we notice thati the
uniform one-way quantum finite automata when accepting with a bounded
error probability can compute only a proper subset of regular sets 
\cite{kowa97}. See also \cite{brpi99} for more recent results on the 
complexity properties of quantum finite automata.
 
\beprop 

For arbitrary Boolean function $f_n$, there exists a read-once
$(2^n,n)$-$\qbp$ that exactly computes $f_n$.

 \eeprop

{\bf Proof:} The proof is evident. The following read-once
$(2^n,n)-\qbp$ $P$ satisfies our proposition. All possible
configurations of $P$ are trivial. That is, a configuration $\qs{\psi}$
of $P$ contains exactly one 1, and all the rest components of
$\qs{\psi}$ are 0.  The initial configuration of $P$ is
$\qs{\psi_0}=(1,0,\dots,0)$.  $P$ reads input variables in order
$x_1,x_2,\dots,x_n$ .

In each step $ i, 1\leq i \leq n $ , $P$ reads input $\sigma_i$ and
transforms its current configuration $\qs{\psi}$ as follows. If
$\sigma_i=0$ then $\qs{\psi}$ does not change.  If
$\sigma_i=1$, then the 1 of the configuration  $\qs{\psi}$ is
``moved'' to $2^{n-i}$ positions to the right in the next
configuration $\qs{\psi'}$.

For an input sequence $\sigma=\sigma_1,\dots.\sigma_n$, denote by 
$l_{\sigma}$ the number of position of 1 in the final (after reading
$\sigma$) configuration of $P$. Clearly we have that $l_{\sigma}\not =
l_{\sigma'}$ iff $\sigma \not = \sigma'$.

Now determine  the set of accepting states $F$ of $P$ as
follows: if $f(\sigma)=1$, then $q_{l_\sigma} \in F$. If $f(\sigma)=0$,
then $q_{l_\sigma} \not \in F$.  
 \Endproof\\

Denote by {\bf EP}\mbox{-}$\qbp_{const}$ the class of
all Boolean functions exactly computed by the constant width and polynomial
length (in the number of function variables) $\qbp$s.

 \beprop
For the complexity class ${\bf NC^1}$ it holds that 

\[{\bf NC^1} \subseteq {\bf EP}\mbox{-}\qbp_{const}.\]
 \eeprop

{\bf Proof:} Proof is evident by a known result of Barrington
\cite{ba89}. Having a permutation deterministic branching program
$P$ of width 5 computing a Boolean function $f_n$ it is easy to construct a
$(const,poly)$-$\qbp$ $P'$ which exactly computes $f_n$. \Endproof 

Consider now the following symmetric Boolean function $MOD_{p_n}$: For an
input $\sigma=\sigma_1,\dots,\sigma_n \in \{0,1\}^n$ we have
$MOD_{p_n}(\sigma)=1$ iff a number of ones in $\sigma$ is divisible by $p_n$,
where $p_n$  is a  prime and $p_n\leq n/2$.

\begin{theorem}\label{qbp-robdd}
 The function $MOD_{p_n}$ can be computed by a read-once ($O(\log
p_n),n$)-$\qbp$  with a one-sided error probability.
 \end{theorem}

The proof of this theorem will be presented in the section below. 
We have clearly that any deterministic OBDD for $MOD_{p_n}$ has
$\Omega(p_n)$ width.

% We clearly  have that any finite deterministic automaton for $L_p$
% needs at least $p$ states. In \cite{amfr98} it was shown that constant
% bounded error finite probabilistic automata also need at least $p$
% number of states to recognize $L_p$. The proof of this lower bound use
% the Markov chain technique (finite probabilistic automata over single
% letter alphabet is exactly a Markov chain). But for probabilistic OBDDs it is
% not the case (probabilistic transitions on the different levels of
% OBDD can be different). Therefore we can not use directly the proof
% method of \cite{amfr98} for general probabilistic OBDD case. But in a
% particular (known enough) case when the transitions (for 0 and 1) in
% each level are the same we can use Markov chain technique for proving
% linear (in $p_n$) width for presentation $MOD_{p_n}$ in such
% probabilistic OBDDs.

\subsection{Proof of Theorem \ref{qbp-robdd} }

The proof is similar to that of \cite{amfr98}. \cite{amfr98}
introduces the following regular set $L_P$. For a prime $p$, a language $L_p$
over a single letter alphabet is defined by $L_p=\{ u: |u|
\mbox{ is divisible by } p\}$. It is proved that for
any $\varepsilon>0$, there is a QFA with $O(\log p)$ states
recognizing $L_p$ with probability $1-\varepsilon$.  \\

 We construct a $\qbp$ $P$ accepting the inputs $\sigma \in
MOD^{-1}_{p_n}(1)$ with the probability 1 and rejecting the inputs $\sigma \in
MOD^{-1}_{p_n}(0)$ with the probability at least 1/8.  Consider a
$(2,n)$-1$\qbp$ $P^k$ for $k \in \{1,\dots,p_n-1\}$. A quantum program
$P^k=\langle T^k,\qs{ \psi_0^k}, F^k\rangle $ is based on the following 2-state quantum
system, $T^k = (\langle i, U^k(0), U^k(1)\rangle )_{i=1}^n$
where \[ U^k(0)= \left(\begin{array}{cc} 1 & 0\\ 0 &1 \\
\end{array}\right), U^k(1)=\left(\begin{array}{cc} \cos(2\pi k/p_n) &
-\sin(2\pi k/p_n)\\ \sin(2\pi k/p_n) & \cos(2\pi k/p_n) \\
\end{array}\right), 
\qs{\psi_0^k}=\left(\begin{array}{c} 1 \\ 0 \\ \end{array}\right), 
 F^k=\{1\}. \]

Denote by $l(\sigma)$ a number  of 1-s in the sequence $\sigma$,
$l(\sigma)= \sum_{i=1}^n \sigma_i$.

\belem[\cite{amfr98}] \label{l-ind}
 After reading an input $\sigma=\sigma_1,\dots,\sigma_n$, the
superposition of $P^k$ is

 \[ \qs{\psi}= \cos \left(\frac{2 \pi \, l(\sigma)k}{p_n}\right)
\qs{1} + \sin \left(\frac {2 \pi \, l(\sigma)k}{p_n}\right) \qs{2}. \]
\eelem

{\bf Proof:} The proof follows from the description of a $\qbp$ $P^k$.

 \Endproof

If the number of ones in an input $\sigma$ is divisible by $p_n$, then 
$2 \pi \, l(\sigma)k/p_n $  is a multiple of $2 \pi$ and
$\cos \Bigl (2 \pi \, l(\sigma)k/p_n \Bigr) =1$,
$\sin \Bigl (2 \pi \, l(\sigma)k/p_n \Bigr) =0$ .  Therefore  all 
$\qbp$s $P^k$  accept   inputs $\sigma \in f^{-1}_{n,p_n}(1)$ with
probability 1. 

Following \cite{amfr98}, we call $P^k$ "good" for input $\sigma \in
MOD^{-1}_{p_n}(0)$ if $P^k$ rejects $\sigma$ with probability at least
$1/2$.
 
\belem
 For  any $\sigma \in MOD^{-1}_{p_n}(0)$, at least $(p_n-1)/2$ of
all  $P^k$ are  ``good''.
\eelem

{\bf Proof:} According to Lemma \ref{l-ind},  after reading an input
$\sigma=\sigma_1,\dots,\sigma_n$ the superposition of $P^k$ is

 \[ \qs{\psi}= \cos \left(\frac{2 \pi \, l(\sigma)k}{p_n}\right)
\qs{1} + \sin \left(\frac {2 \pi \, l(\sigma)k}{p_n}\right) \qs{2}. \]

Therefore, the probability of accepting the input $\sigma \in
MOD^{-1}_{p_n}(0)$ is $\cos^2 \Bigl( 2 \pi \, l(\sigma)k/p_n\Bigr)$.
$\cos^2 \Bigl( 2 \pi \, l(\sigma)k/p_n\Bigr)\leq 1/2$ iff $\left|\cos
\Bigl( 2 \pi \, l(\sigma)k/p_n\Bigr)\right|\leq 1/\sqrt{2}$.  This
happens if and only if $\Bigl( 2 \pi \, l(\sigma)k/p_n \Bigr)$ is in
$[\pi/4+2\pi j, 3\pi/4 + 2\pi j]$ or in $[5\pi/4+2\pi j, 7\pi/4 + 2\pi
j]$ for some $j \in N$.  $\Bigl( 2 \pi \, l(\sigma)k/p_n \Bigr) \in
[\pi/4+2\pi j, 3\pi/4 + 2\pi j]$ iff $\Bigl( 2 \pi \, (l(\sigma)k
\bmod{p_n})/p_n\Bigr)\in [\pi/4, 3\pi/4 ]$.  $p_n$ is prime, and
$l(\sigma)$ is relatively prime with $p_n$. Therefore, $l(\sigma)
\bmod{p_n}$, $2l(\sigma) \bmod{p_n}, \dots,$ $ (p_n-1)l(\sigma)
\bmod{p_n} $ is $1,2,\dots,p_n-1$ in different order. Consequently, it
is enough to find the power of a set $I=\{i_1,\dots, i_l\}\subset
\{1,\dots, p_n-1\}$ such that $ 2 \pi \, i_j \, /p_n \in [\pi/4,
3\pi/4]$ or $ 2 \pi \, i_j \, /p_n \in [5\pi/4, 7\pi/4]$.  Since $p_n$
points $2\pi/p_n,\dots, 2\pi(p_n-1)/p_n, 2\pi$ are regularly distributed on a
circumference, and sectors $[\pi/4,3\pi/4 ]$ and $[5\pi/4,7\pi/4]$ are
exactly a half of the circumference, we have  $|I|\geq \lfloor p_n/2
\rfloor \geq (p_n-1)/2.$

 \Endproof

Following   \cite{amfr98}, we  call   a set  of quantum programs $S=\{P^{i_1},
\dots, P^{i_t}\}$    ``good'' for
$\sigma \in f^{-1}_{n,p_n}(0)$
if at least $1/4$ of all its elements are ``good'' for this
$\sigma$.

\belem 
There is a set $S$ of 1$\qbp$s  with  $|S|=t=\ceiling{16 \ln p_n}$ which is
``good'' for all inputs $\sigma \in MOD^{-1}_{p_n}(0).$
\eelem

{\bf Proof:}
 We consider the following procedure ${\cal A}$ for a 
construction of a set $S$. 

\begin{quote}
 For a fixed input $\sigma$ with $l(\sigma)
\leq p_n-1, $ ${\cal A}$ selects a quantum branching program uniformly at
random from $\{P^1,\dots, P^{p_n-1}\} .$
\end{quote}

 The probability of selecting a ``good'' $\qbp$ at each step is at least
1/2. Using Chernoff inequality, we have that the probability that less than
1/4 fraction of all $\qbp$s from the set $S$ are ``good'', for any fixed
$\sigma$ with $l(\sigma)\leq p_n-1$, is at most

\[ \exp((-16 \ln p_n)/2 (1/2)^2/2=1/p_n.\]

Hence the probability that a constructed set is not ``good'' for at
least one input $\sigma$ with $l(\sigma)\leq p_n-1$, is at most
$(p_n-1)/p_n>0$. Therefore there exists a set which is ``good'' for all
inputs $\sigma$ with $l(\sigma)\leq p_n-1.$ This set is ``good'' for the
inputs $\sigma$ with $l(\sigma)> p_n$ as well, since any $\qbp$ $P^k$
returns a current superposition of a starting superposition
after reading all $p_n$ ones, and hence it works the same way on inputs
$\sigma, \sigma'$ with $l(\sigma)=l(\sigma')\bmod p_n. $

 \Endproof

We construct a 1$\qbp$ $P$ accepting inputs $\sigma \in
MOD^{-1}_{p_n}(1)$ with probability 1 and rejecting inputs $\sigma \in
MOD^{-1}_{p_n}(0)$ with probability at least 1/8 as follows. A 1$\qbp$
$P$ consists of $\qbp$s from ``good'' set $S=\{P^{i_1}, \dots,
P^{{i_{t}}}\}$, which work in parallel. In the starting
superposition of $P$ all these programs have equal
amplitudes.

The inputs $\sigma \in MOD^{-1}_{p_n}(1)$ are always accepted with the
probability 1 because all the $P^k$s accept them. For any input $\sigma \in
MOD^{-1}_{p_n}(0)$ at least 1/4 of all $P^k \in S$ reject it with
probability at least 1/2, and the total probability of rejecting any
$\sigma \in MOD^{-1}_{p_n}(0)$ is at least 1/8.

We can make the error as small as possible using a standard technique
for reducing an error probability for a one-sided error computation. That is, we take
$d=d(\varepsilon)$ copies of such a 1$\qbp$ $P$ and run them uniformly at
random.  In this case the width of 1$\qbp$ will be $O(\log p_n)$.
\Endproof

\begin{itemize}

\item We call a branching program $P$ {\em stable} if its transformations
do not depend on the level of $P$.

\end{itemize}

From the proof of our theorem we have that the constructed $\qbp$ for
$MOD_{p_n}$ is a stable branching program.

\becor The function $MOD_{p_n}$ can be computed by a stable read-once
 ($O(\log p_n),n$)-$\qbp$ with a one-sided error. 
 \eecor

Below we show that $MOD_{p_n}$ function is hard, in fact, for the randomized OBDDs.

\subsection{Lower Bound for Randomized OBDDs for $MOD$}

Randomized OBDDs were introduced and firstly investigated in
\cite{ak98}, see also \cite{we00}.

 \bethm

Any stable probabilistic  OBDD computing $MOD_{p_n}$ has width at least 
$p_n$.
\eethm

{\bf Proof:} Assume that there is a stable probabilistic OBDD $P$ of
width $q < p_n$ computing $MOD_{p_n}$ with probability
$1/2+\varepsilon$ for a fixed $\varepsilon\in (0,1/2]$. We can assume without loss of generality that   
each level of $P$ has exactly $q$ nodes.  
Let $\mu^j=(\mu^j_1,\dots,\mu^j_q)$ be a probability
distribution of states of $P$ on the $j$-th level, where $\mu^j_i$ is the 
probability of being in the $i$-th node of the $j$-th level. We can
describe a computation of $P$ on an input $\tilde
\sigma=\sigma_1,\dots,\sigma_n$ as follows:
 
\begin{itemize}

\item A computation of $P$ starts from the initial probability
distributions vector $\mu^0$.  

\item At the $j$-th step, $1\leq j \leq n$, $P$ reads an input $\sigma_{i_j}$
and transforms a vector $\mu^{j-1}$ to $\mu^j=\mu^{j-1} A$, where $A$ is a $q
\times q$ stochastic matrix, $A=A(0)$ if $\sigma_{i_j}=0$ and $A=A(1)$ if
$\sigma_{j_k}=1$.

\item After the last ($n$-th) step of the computation, $P$ accepts the input
$\tilde \sigma$ with probability $P_{acc}(\tilde \sigma)=\sum_{i \in
F}\mu_i$. If $f(\tilde \sigma)=1$, then we have $P_{acc}(\tilde
\sigma)\geq 1/2 + \varepsilon$, else we have  $P_{acc}(\tilde
\sigma)\leq 1/2 - \varepsilon$.

\end{itemize}

We assume without loss of generality that $P$ reads inputs in the natural order $x_1,\dots,x_n$.
We consider all inputs $\tilde
\sigma_n,\dots,\tilde \sigma_1$, such that
$\tilde\sigma_i=\tilde\sigma_i^0 \tilde\sigma_i^1,$ where
$\tilde\sigma_i^0=\underbrace{0 \dots 0}_{n-i},$
$\tilde\sigma_i^1=\underbrace{1 \dots 1}_{i}.$

For $i \in \{1,\dots, n\}$ we denote by $\mu^i$ a probability distribution
after reading the part $\tilde \sigma_i^0$. That is, $\mu^i=\mu^0
A(0)\cdots A(0)=\mu^0 A^{n-i}(0)$.  There are only ones in the $\tilde
\sigma_i^1$, hence a computation after reading $\tilde \sigma_i^0$ can
be described by a Markov chain. In this case $\mu^i$ is the initial
probability distribution for a Markov process, $A(1)$ is the
transition probability matrix. 
The states
of this Markov chain are either ergodic or transient cf., e.g., \cite{kesn60}. An
ergodic set of states is a set which a process cannot leave if it once
entered. A transient set of states is a set which a process can leave,
but cannot return if it once left. An ergodic state is an element of
an ergodic set. A transient state is an element of transient set.

An arbitrary Markov chain $C$ has at least one ergodic set. $C$ can be a
Markov chain without any transient set. If a Markov chain $C$ has more than
one ergodic set, then there is no interaction between these
sets. Hence we have two or more unrelated Markov chains lumped together.
Those chains can be studied separately. If a Markov chain consist of a single
ergodic set, then the chain is called an ergodic chain. According to the usual
classification, every ergodic chain is either regular or cyclic.

If an ergodic chain is regular, then a sufficiently high power of a state
transition matrix $A$ has only positive elements. Thus no matter where
the process starts, after a sufficient number of steps it can be in
any state. Moreover, there is a limiting vector of probabilities of
being in the states of the chain which do not dependent on the initial state.

 If a Markov chain is cyclic, then a chain has a period $t$ and all
its states are subdivided into $t$ cyclic subsets $(t>1)$. For a given
starting state a process moves through the cyclic subsets in a
definite order, returning to the subset with the starting state after
every $t$ steps. It is known that after the sufficient time elapsed,
the process can be in any state of a cyclic subset appropriate at 
the given moment. Hence for each of $t$ cyclic subsets, the $t$-th power of
a state transition matrix $A^t$ describes a regular Markov chain.
Moreover, if an ergodic chain is a cyclic chain with a period $t$, it
has at least $t$ states.

From the assumption $q < p_n$ in the proof, we get that $t < p_n$ for
every cyclic chain. We denote by $D$ the least common multiply of all
such $t$. Because $p_n$ is prime, $t$ is relatively prime to $p_n$,
$D$ is relatively prime to $p_n$, and so is any positive
power $D^m$ of $D$.

 Let $\alpha^k$ be a probability distribution after reading the part
 $\tilde \sigma_k^1$ of $\tilde\sigma_k.$ That is, $\alpha^k=\mu^k A^k
 (1).$ We can assume that there is a single accepting state, without
 loss of generality. Let $\alpha^k_{acc}$ be the probability of being in 
 accepting state after reading the input $\tilde\sigma^1_k.$ Since
 after every $D$ steps a process can be in any state comprising an  
 accepting state, $D$-th power of matrix $A$ describes a regular
 Markov chain for that set. From the theory of Markov chains we have
 that there exists $\alpha_{acc}$ that $\lim_{k \to
 \infty}\alpha^{kD}_{acc}=\alpha_{acc}$. Hence for any $\varepsilon>0$
 it holds that
 \[|\alpha^{D^m}_{acc}-\alpha^{D^mp_n}_{acc}|<2\epsilon\] for $m$
 large enough. As $P$ ($1/2+\varepsilon$)-computes $MOD_{p_n}$, we have that
 $\alpha^{D^m p_n}_{acc}\geq 1/2+\varepsilon$ and
 $\alpha^{D^m}_{acc}\leq 1/2-\varepsilon$, a contradiction with the
 inequality above.

%herefore the probability distributions vector

% Let's $\alpha_k$ is a
%probability being in accepting state after $k$ steps. Since a
%probability distributions vector strives for limiting vector of
%probabilities, $\alpha_k$ strives for $\alpha$ when $k\rightarrow
%\infty$. Therefore for a fixed $\epsilon>0$ there is $k$, that it is
%hold $|\alpha_k-\alpha_{k+i}|<2\varepsilon$ for any $i\in N$.

%Since $f(\tilde\sigma_{D^m})=0$ but $f(\tilde \sigma_{D^m{p_n}})=1$,
%the total of the probability to be in the accepting state must be less
%than $(1/2-\epsilon)$ for $\tilde\sigma_D^1$ and greater than
%$(1/2+\epsilon)$ for $\tilde \epsilon_{D{p_n}}^1$, not depended of the
%initial state. Contradiction.

 \Endproof

%**********************************************

\section{Lower Bounds}

Below we present a general lower bound on the width of 1$\qbp$s 
and compare it with the width of deterministic OBDDs computing the same
function.

 \begin{theorem}\label{lb1}
 Let $\varepsilon\in(0,1/2)$. Let $f_n$ be a Boolean function which is
 $(1/2+\varepsilon)$-computed (computed with a margin $\varepsilon$) by a
 1$\qbp$ $Q$. Then it holds that

 \[ width(Q)=\Omega\left(\frac{\log width(P)}{\log\log
 width(P)}\right) \]
 where $P$  is a deterministic OBDD of minimal width which computes 
 $f_n$.

 \end{theorem}

The next theorem presents a more precise lower bound for a particular margin
$\varepsilon$ of computation.

\begin{theorem} \label{lb2}
 Let $\varepsilon\in(3/8,1/2)$. Let $f_n$ be a Boolean function which is
 $(1/2+\varepsilon)$-computed (computed with a margin $\varepsilon$) by a
 1$\qbp$ $Q$. Then it holds that

 \[ width(Q)=\Omega\left(\frac{\log width(P)}{2\log
 (1+1/\tau)}\right) \]
 where $P$  is a deterministic OBDD of minimal width computing
 $f_n$ and \\
 $\tau=\sqrt{1+2\varepsilon-4\sqrt{1/2-\varepsilon}}.$

 \end{theorem}

Proofs of the above theorems are presented in the section below.

\subsection{Proofs of Theorems \ref{lb1} and \ref{lb2}}

Proofs of Theorems \ref{lb1}, and \ref{lb2} use a similar idea. We
construct a deterministic OBDD $P$  that computes 
the same function  $f_n$ and

\begin{equation}\label{ubB}
width(P) \leq \left( 1+\frac{2}{\theta}\right)^{2 \, width({Q})}.
\end{equation}

Proofs of Theorems \ref{lb1}, \ref{lb2} differ only in an estimation of a 
parameter $\theta >0$ depending on $\varepsilon$.

\subsubsection{A Deterministic OBDD-Presentation of a 1$\qbp$}

 Let $d=width({Q})$.  Let $\pi=\{i_1,i_2,\dots, i_n\}$ be an ordering
of testing variables of $Q$.  From now on we assume that the input
sequences $\sigma\in\{0,1\}^n$ are ordered by an order $\pi$
determined by $Q$.  We define a deterministic OBDD $L\!Q$ based on $Q$
as follows. $LQ$ use the ordering $\pi$ of testing variables 
represented by the following labeled complete binary tree.

\begin{itemize}
 \item The initial node of $L\!Q$ is marked by an initial configuration
$\qs{\psi_0}$ of $Q$. Two outgoing vertices of the initial node are marked
by $x_{i_1}=1$ and $x_{i_1}=0$.

 \item Two nodes of $L\!Q$ on the level 1 are marked by the configurations
$\qs{\psi_1(0)}$ and $\qs{\psi_1(1)}$ of $Q$ where
$\qs{\psi_{1}(\sigma_1)}$ is the configuration after the first step of
computation after reading $x_{i_1}=\sigma_1$ for $\sigma_1\in\{0,1\}$.

A vertex $x_{i_1}=\sigma_1$ leads from the node $\qs{\psi_0}$ to the
node $\qs{\psi_{1}(\sigma_1)}$ iff $\qs{\psi_{1}(\sigma_1)}
=U_1(\sigma_1)\qs{\psi_0}$.

\item Consider a level $j$ of $LQ$.  $2^j$ nodes of $L\!Q$ of the
level $j$ are marked by the configurations $\{
\qs{\psi_j(\sigma_1\dots\sigma_j)}\in\Psi : \sigma_1\dots\sigma_j
\in\{0,1\}^j \}$ where $\qs{\psi_j(\sigma_1\dots\sigma_j)}$ is a
configuration of $Q$ after reading the first part
$\sigma_1\dots\sigma_j$ of the input $\sigma\in\{0,1\}^n$.

A vertex (marked by $x_{i_{j+1}}=\sigma_{j+1}$) from the node
$\qs{\psi_j(\sigma_1\dots\sigma_j)}$ leads to the node
$\qs{\psi_{j+1}(\sigma_1\dots\sigma_j\sigma_{j+1})}$ iff
$\qs{\psi_{j+1}(\sigma_1\dots\sigma_j\sigma_{j+1})} =
U_{j+1}(\sigma_{j+1})\qs{\psi_j(\sigma_1\dots\sigma_j)}$.

\item Consider the last level $n$ of $LQ$. We mark $2^n$ nodes of $L\!Q$
on the level $n$ by the configurations
$\qs{\psi_n(\sigma_1\dots\sigma_n)}\in\Psi :
(\sigma_1\dots\sigma_n)\in\{0,1\}^n$ and in addition we mark them by 0
and 1 as follows. We mark node $\qs{\psi_n(\sigma_1\dots\sigma_n)}$ by
1 if for configuration $\qs{\psi_n(\sigma_1\dots\sigma_n)}$ it
holds that $p_{accept}(\sigma_1\dots\sigma_n)\geq 1/2+\varepsilon$.
We mark a node $\qs{\psi_n(\sigma_1\dots\sigma_n)}$ by 0 if for
configuration $\qs{\psi_n(\sigma_1\dots\sigma_n)}$ it  holds that
$p_{accept}(\sigma_1\dots\sigma_n)\leq 1/2-\varepsilon$.
 
 \end{itemize}

\beprop
A deterministic OBDD $LQ$ computes the same Boolean function
$f_n$ as $Q$. 
\eeprop

{\bf Proof:} Evident and follows from the construction of
$LQ$. \Endproof

\subsubsection{A Metric Automaton Characterization of $LQ$}

We  view now an OBDD $LQ$ with an ordering $\pi$ of testing variables 
as the following metric time-variant automaton that reads its input
sequences $\sigma\in\{0,1\}^n$ in an order $\pi$:

\[ LQ=\langle \{0,1\}, \Psi, \{\delta_j\}_{j=1}^n, \qs{\psi_0}, {\cal
F}_\varepsilon \rangle \]
 where $\{0,1\}$ is the input alphabet, $\Psi=\{\qs{\psi} \}$ is a
set of states (set of all possible configurations of $Q$ during its
computations on inputs from $\{0,1\}^n$). That is,
$\Psi=\cup_{j=0}^n\Psi_j$ where $\Psi_j$ is a set of states of $LQ$
on the level $j$. An automaton transition function $\delta_j :
\Psi_{j-1}\times \{0,1\} \to \Psi_{j}$ determines transitions in the
step $j$, $1\leq j\leq n$, ($\delta_j$ is defined according to the 
transitions of $LQ$ in the level $j-1$).  Finally $\qs{\psi_0}$ is the
initial state and ${\cal F}_\varepsilon =\{\qs{\psi}\in \Psi_n :
||M\qs{\psi}||^2\geq 1/2+\varepsilon\}$ is the accepting set of states
of $LQ$.

For $j\in \{1,\dots, n\}$, we  denote by $\Delta_j : \Psi_{j-1}\times
\{0,1\}^{n-j+1} \to \Psi_n$  the automaton transitive closure of the
sequence $\delta_j,\dots,\delta_n$ of the transition functions. That is,

\[ \Delta_j(\qs{\psi}, \sigma_j\dots\sigma_n)=\delta_n (\dots
(\delta_j(\qs{\psi},\sigma_j),\dots,\sigma_n). \]

\belem\label{t-lb} 
 Let $f_n$ be a Boolean function ($1/2+\varepsilon$)-computed by
a ${LQ}$.  Let $\theta >0$, and for arbitrary $\qs{\psi}\in{\cal
F}_\varepsilon$ and arbitrary $\qs{\psi}'\in\Psi_n\backslash {\cal
F}_\varepsilon$ it holds that

% \begin{equation}\label{t1}
\[ || \, \qs{\psi} -\qs{\psi'} \, || \geq \theta. \]
% \end{equation}

Then, there exists a deterministic OBDD $B$ which
computes  $f_n$ and 
 \[width(B) \leq \left(1+ \frac{2}{\theta}\right)^{2d}. \]
 
\eelem

{\bf Proof:} We recall first some known notions concerning metric spaces 
(see \cite{al}). A Hilbert space ${\cal H}_d$ is a
metric space with a metric defined by the norm $||\cdot ||$. The points
$\mu, \mu'$ from ${\cal H}_d$ are connected through a $\theta$-chain
if there exists a finite set of points $\mu_1, \mu_2, \ldots, \mu_m$
from ${\cal H}_d$ such that $\mu_1=\mu, \mu_m=\mu'$ and
$||\mu_{i}-\mu_{i+1}||\leq\theta$ for $i\in\{1,\ldots,m-1\}$. A subset $C$ of ${\cal H}_d$
is called a $\theta$-component if arbitrary
two points $\mu, \mu'\in {\cal C}$ are connected through
a $\theta$-chain. It is known \cite{al} that if ${\cal D}$ is a finite
diameter subset of a subspace of ${\cal H}_d$ (a diameter of ${\cal D}$ is
defined as $\sup_{\mu,\mu'\in {\cal D}}\{||\mu-\mu'||\}$ then for
$\theta > 0$ ${\cal D}$ is partitioned to a finite number $t$ of its
$\theta$-components.

A set $\Psi$ of states of $L\!Q$ belongs to the sphere
of radius 1 which has center $(0,0,\dots,0)$ in ${\cal H}_d$ because
for all $\qs{\psi}\in\Psi$ it holds that $||\, \qs{\psi}\, ||=1$.  For
each $j\in\{0,\dots, n\}$, denote by $[\Psi_j]=\{C_1,\dots,C_{t_j}\}$
the set of $\theta$-components of $\Psi_j\subset {\cal H}_d$.

From the condition of our Lemma it follows that a subset ${\cal
F}_\varepsilon $ of $\Psi_n$ is a union of some $\theta$-components of
$\Psi_n$. The transition functions $\delta_j$, $1\leq j\leq n$, preserve
the distance. That is, for arbitrary $\qs{\psi}$ and $\qs{\xi}$ from $\Psi_j$
and arbitrary $\gamma\in\{0,1\}$, it holds that

 \begin{equation}\label{dist}
 ||\, \qs{\psi} - \qs{\xi} \, ||=||\delta_j(\qs{\psi},\gamma)
-\delta_j(\qs{\xi},\gamma)||.
 \end{equation}

From (\ref{dist}) we have that for $C\in[\Psi_j]$ and for
$\gamma\in\{0,1\}$ there exists $C'\in[\Psi_{j+1}]$ such
that $\delta_j(C,\gamma) = C'$.  Here $\delta_j(C,\gamma)$ is defined as
$\delta_j(C,\gamma) =\cup_{\qs{\psi}\in C}\delta_j(\qs{\psi},\gamma)$.

Now we describe a deterministic OBDD $B$ in terms of a time-variant finite
automaton that computes  $f_n$.

\[ B=\langle \{0,1\}, \, [\Psi], \, \{\delta_j\}_{j=1}^{n}, \,
C_0, \, F \rangle \]
 where $[\Psi]=\cup_{j=0}^{n} [\Psi_j]$ is a set of states of $B$
($[\Psi_j]$ is a set of states on the step $j$ of a
computation of $B$); \\
$\delta_j :[\Psi_{j-1}]\times \{0,1\}\to [\Psi_j]$ is a transition
function of $B$ in the step $j$;  \\
an initial state $C_0=\{\qs{\psi_0}\}$ is a one-element
$\theta$-component of $\Psi_0$; we define $F$ by  
$F=\{ C_i\in [\Psi_n] : C_i\subseteq {\cal
F}_\varepsilon \}$.

From our construction we have that $B$ and ${LQ}$ compute
the same function $f_n$.  The width of $B$ is $t=\max
\{t_0,\dots,t_n\}$.

Let $t=t_j$
We estimate a number $t$ of
$\theta$-components (number of states of $B$) of $\Psi_j$ as follows.
For each $\theta$-component $C$, we  select one point $\qs{\psi}\in C$. If we
draw a sphere of the radius $\theta/2$ with the center $\qs{\psi}\in C$
then all such spheres do not intersect pairwise. All these spheres (t many)
are in a larger sphere of radius $1+\theta/2$ which has center
$(0,0,\dots,0)$. The volume of a sphere of radius $r$ in ${\cal
H}_d$ is $cr^{2d}$, where the constant $c$ depends on a metric of
${\cal H}_d$. Note that for estimating the volume of the sphere we
should take into account that ${\cal H}_d$ is a $d$-dimensional {\em
complex} space and each complex point is a 2-dimensional point.  So it
holds that

 \[ width(B) \leq \frac{c\left(1+\theta/2\right)^{2d}}{
c\left(\theta/2\right) ^{2d}}
= \left(1+ \frac{2}{\theta}\right)^{2d}. \]
\Endproof

Below we formulate a technical lemma that estimates a 
number of components of $\Psi$ for different $\varepsilon$.

\belem\label{dif-the}
 Let an $LQ$ ($1/2+\varepsilon$)-computes a function $f_n$. Then for
arbitrary $\qs{\psi}\in{\cal F}_\varepsilon$ and arbitrary
$\qs{\psi'}\not\in{\cal F}_\varepsilon$ it holds that

 \begin{enumerate}\label{t1}
 \item $||\, \qs{\psi} -\qs{\psi'} \,|| \geq \theta_1=\varepsilon/\sqrt{d}$

and

 \item
 $ ||\, \qs{\psi} -\qs{\psi'} \,|| \geq
\theta_2=\sqrt{1+2\varepsilon-4\sqrt{1/2-\varepsilon}}$.  
 \end{enumerate}

\eelem

{\bf Proof:} For the sake of simplification we denote a configuration
$\qs{\psi}= z_1\qs{1} +\cdots +z_d\qs{d}$ just by $\psi=(z_1,\dots,
z_d)$.  Let $\psi=(z_1,\dots, z_d)$ and $\psi'=(z'_1,\dots z'_d)$.
Consider a norm $||.||_1$ defined as $||\psi||_1= \sum_{i=1}^d|z_i|$.
 
{\bf 1.} From the definition of ${ LQ}$ it holds
that

\[ 2\varepsilon \le \sum_{s_i \in F}{(|z_i|^2-|z'_i|^2)}=
\sum_{s_i \in F}{(|z_i|-|z'_i|)(|z_i| + |z'_i|)}\le \]

\[ \le  2\sum_{s_i \in F}{(|z_i|-|z'_i|)}\le 2\sum_{s_i \in F}{|z_i-z'_i|}
\le 2||\psi-\psi'||_1 \]

Using an inequality  
\begin{equation}\label{ineq}
a_1 b_1 + a_2 b_2 +...+a_d b_d\le
\sqrt{a_1^2+a_2^2+...+a_d^2} \sqrt{b_1^2+b_2^2+...+b_d^2},
\end{equation}

for  $b_1=b_2=\ldots =b_d=1$ we get that 
 $||\psi||_1 \le \sqrt{d}||\psi||$. Therefore,

\[ 2\varepsilon \le 2||\psi-\psi'||_1 \le 2\sqrt{d}||\psi-\psi'|| \]

 Finally, we have
\[ ||\psi-\psi'|| \ge \varepsilon/\sqrt{d}. \]

{\bf 2.} Consider now the next variant of a lower bound for $||\psi-\psi'||$. 

\begin{eqnarray}
 ||\psi-\psi'|| &=& \sqrt{\sum_{i=1}^d{|z_i-z'_i|^2}}\ge
\sqrt{\sum_{i=1}^d{(|z_i|-|z'_i|)^2}}=  \nonumber \\
        &=& \sqrt{\sum_{i=1}^d{|z_i|}^2 + 
\sum_{i=1}^d{|z'_i|}^2 - 2\sum_{i=1}^d{|z_i| |z'_i|}}\ge  \nonumber \\
      &\ge& \sqrt{\sum_{s_i \in F}{|z_i|^2}+
\sum_{s_i\not\in F}{|z'_i|^2-2\sum_{s_i\in F}{|z_i| |z'_i|}}-
2\sum_{s_i\not\in F}{|z_i| |z'_i|}}. \nonumber 
 \end{eqnarray}

From the definition of an ${LQ}$ we have that $\sum_{s_i\in
F}{|z_i|^2}\ge 1/2+\varepsilon$, $\sum_{s_i\not \in F}{|z'_i|}^2\ge
1/2+\varepsilon$. Now, from the above we get that

\[ ||\psi-\psi'|| \geq 
\sqrt{1/2+\varepsilon+1/2+\varepsilon-2\sum_{s_i\in F}{|z_i| |z'_i|}-
2\sum_{s_i\not \in F}{|z_i| |z'_i|}}. \]

Using inequality (\ref{ineq}) we get 

\[ ||\psi-\psi'|| \geq 
 \sqrt{1+2\varepsilon-2\sqrt{\sum_{s_i \in F}{|z_i|^2}} 
\sqrt{\sum_{s_i\in F}{|z'_i|^2}}-2\sqrt{\sum_{s_i \not \in F}{|z_i|^2}}
\sqrt{\sum_{s_i \not\in F}{|z'_i|^2}}}. \]

Using  the property 
 $\sum_{s_i \not\in F}{|z_i|^2}\le 1/2-\varepsilon$,
 $\sum_{s_i \in F}{|z'_i|^2}\le 1/2-\varepsilon$,
 $\sum_{s_i}{|z_i|^2}\le 1$, and $\sum_{s_i}{|z'_i|^2}\le 1$,
 we finally get that

\[ ||\psi-\psi'|| \geq 
\sqrt{1+2\varepsilon-4\sqrt{1/2-\varepsilon}}=\theta_2. \]
\Endproof\\

Note that the  lower bound  above for $||\psi-\psi'||$ is
nontrivial (positive) if $\varepsilon\in (\alpha,1/2)$ where $\alpha$
is about 3/8. For $\varepsilon \in (0,\alpha]$ it holds that
$1+2\varepsilon-4\sqrt{1/2-\varepsilon} \leq 0$. 
In that case the  lower bound 
$||\psi-\psi'||\geq \varepsilon/\sqrt{d}$ is more precise. \\

Now we turn to the formal estimations  of the lower bounds of Theorems
\ref{lb1} and \ref{lb2}. \\

{\bf Proof of Theorem \ref{lb1}:} From Lemma \ref{t-lb} and Lemma
\ref{dif-the}, it follows that 

\[ t\leq \left(1+\frac{2\sqrt{d}}{\varepsilon}\right)^{2d} \]

 or $\log t =O(d\log d)$. From that we get that

 \[ d =\Omega\left(\frac{\log t}{\log\log t}\right). \]

{\bf Proof of Theorem \ref{lb2}:} From Lemma \ref{t-lb} and Lemma
\ref{dif-the}, it follows that  

 \[ t \leq \left(1+ \frac{2}{\theta_2}\right)^{2d} \]
or $2d \geq  \log t / \log
(1+2/\theta_2)$. From this we have that

\[ d \geq \frac{\log t} {2\log (1+1/\theta_2)}. \]

 \Endproof

\end{document}